\documentclass[12pt,preprint]{aastex}
\shorttitle{}
\shortauthors{}

\begin{document}

\title{
Relationship between different channel light curves of gamma-ray
burst pulses shown in aspects other than the pulse width }

\author{Yi-Ping Qin\altaffilmark{1,2,3}}

\altaffiltext{1}{National Astronomical Observatories/Yunnan
Observatory, Chinese Academy of Sciences, P. O. Box 110, Kunming,
Yunnan, 650011, P. R. China}

\altaffiltext{2}{Physics Department, Guangxi University, Nanning,
Guangxi 530004, P. R. China}

\altaffiltext{3}{E-mail:ypqin@public.km.yn.cn}

\begin{abstract}
The $FWHM$ width is a well known quantity representing the characteristics
of pulses of gamma-ray bursts (GRBs), for which many studies were presented.
However, there are some other elements which can describe other aspects of
the pulses that the width cannot. In this paper, we employ the peak count
rate{\bf \ }$C_p$ and the total count $C_{total}$ of light curves to study
in the corresponding aspects the relationship between different channel
light curves. To make a direct comparison between count rates of different
channel light curves we introduce a plot of $C(\tau )$ versus $C_H(\tau )$,
where $C(\tau )$ is the count rate of a channel and $C_H(\tau )$ is the
count rate of a definite cannel, channel H (see the text). According to the
plot we define $\Delta C_{\max }$ as the maximum deviation of the two count
rate values of $C(\tau )$ associated with a same count rate value of $%
C_H(\tau )$ and define $\Delta S$ as the area confined by the
close curve of $C(\tau )$ in the plot to measure the difference of
the rising and decaying portions of a light curve relative to the
count rate of channel H. Under the assumption that some GRBs
observed are in the stage of fireballs which expand
relativistically, predictions on the relationships between the
four quantities{\bf \ (}$C_p$, $C_{total}$, $\Delta C_{\max }$,
and $\Delta S$) and energy within a wide band, calculated with
different rest frame radiation forms and two typical Lorentz
factors ($\Gamma =20$ and $200$), are made and presented, which
would make the test of our model with the coming Swift data
easier. Interpretations to the relationships within the mechanism
of fireballs are also presented. \end{abstract}


\section{Introduction}

Observation reveals that some simple gamma-ray bursts (GRBs) with
well-separated structure consist of fundamental pulses which are seen to
comprise a fast rise and an exponential decay (FRED) phases (see, e.g.,
Fishman et al. 1994). The FRED pulses together with the hardness-intensity
correlation were interpreted as signatures of the relativistic curvature
effect (the Doppler effect over the fireball surface) (Fenimore et al. 1996;
Ryde and Petrosian 2002; Kocevski et al. 2003; Qin et al. 2004, hereafter
Paper I). Accounting for the great output rate of radiation of GRBs, a
fireball model was proposed as early as in 1980s (Goodman 1986; Paczynski
1986), and based on it, the shifting of the spectrum or the boosting of the
radiation of the objects and other phenomena, due to the Doppler effect over
the fireball surface, could be expected (see, e.g., Krolik and Pier 1991;
Meszaros and Rees 1998; Hailey et al. 1999; Qin 2002, 2003).

One could find from the BATSE data that FRED pulse light curves of GRBs in
different energy channels differ significantly in the magnitude and the
width of the curves. However, from these light curves a self-similarity
across energy bands could be observed (see, e.g., Norris et al. 1996). In
the past few years, many attempts of interpretation of the light curves
observed have been made (see, e.g., Fenimore et al. 1996; Norris et al.
1996; Norris et al. 2000; Ryde and Petrosian 2002; Kocevski et al. 2003).
Statistical analysis based on the fit to the light curve revealed that the
temporal scale factors of a given pulse measured at different energies are
related to the corresponding energies by a power law (Fenimore et al. 1995;
Norris et al. 1996; Nemiroff 2000), which was suspected to result from a
relative projected speed or a relative beaming angle (Nemiroff 2000). Paper
I shows that the power law relationship between the pulse width and energy
as a consequence of the Doppler effect over the fireball surface is due to
different active areas of the fireball surface corresponding to the majority
of photons of the channels observed. An investigation on the relationship
was performed in detail previously and based on it the expected behavior of
the relationship in the Swift band was presented (see Qin 2004). However,
the width of pulses reflects only one aspect of the light curves. Generally,
for a certain event, if only a relationship is observed, one might find
several mechanisms accounting for it; however, when several relationships
are observed, the number of interpretation might be fewer or even unique. It
is understood that, generally, the more relationships observed, the few
number of possible interpretations. Therefore, to have the model completely
checked or to sufficiently understand how different channel light curves are
related, other quantities rather than the pulse width such as the magnitude
or the total count should also be studied.

Shown in Fig. 10 of Paper I are the light curves of the four BATSE channels
of GRB 951019, where the fitting curves of the four channels share the same
formula with the same parameters, differing only in the energy integral
range. In this analysis, no parameters other than the energy ranges are
allowed to vary with the energy channel concerned. However, even with this
figure, the following question cannot easily be answered: for the same count
rate of a channel measured at different times, what would happen to the
count rates of other channels measured at these times? This suggests that,
to show the relation between different channel light curves, figures other
than this one might also be at work and might work better in some particular
aspects.

In the following, we will study in detail the relation between different
channel light curves with some quantities other than the width, and based on
the analysis predictions on the behavior of the corresponding relationships
in the Swift band will be made.

\section{Characters of light curves determined by the energy range of
channels}

Through out this paper, the relation between different channel light curves
of GRB pulses is studied under the assumption that the sources concerned are
in the stage of fireballs which expand relativistically and isotropically.
Here we study how some characters of light curves are related with the
corresponding energy.

The width of pulses is a well-known character of light curves determined by
energy channels. Let us consider two other quantities which are also
characters of light curves associated with the corresponding channels. One
is the peak count rate and the other is the total count. The distributions
of these two quantities for both long and short GRBs were investigated
previously (McBreen et al. 2001, 2003). Here, we pay our attention to the
relations of these quantities with energy. The formula for calculating the
count rate of a fireball source is (A1) (see Appendix A). In this paper, we
consider only the cases of local Gaussian pulse (A2) and local power law
pulse (A3) (see Appendix A). Two kinds of fireball bursts are concerned. One
is a typical hard burst with $\Gamma =200$ and the other is a typical soft
burst with $\Gamma =20$ and they are assumed to share the same rest frame
radiation spectrum and the same local pulse (see Qin 2004). It is known that
the peak energy $E_p$ could serve as a quantity describing the hardness of
the spectrum of a burst observed (see Ford et al. 1995). In the case of a
rest frame Band function (Band et al. 1993) with $\alpha _0=-1$ and $\beta
_0=-2.25$, when adopting $\nu _{0,p}=0.75keVh^{-1}$ we find that the typical
hard burst has a peak energy at $E_p=250keV$ and the typical soft burst is
associated with $E_p=25keV$ (see Qin 2002 Table 4).

\subsection{The case of a typical rest frame Band function spectrum}

Here we adopt a typical Band function spectrum with $\alpha _0=-1$ and $%
\beta _0=-2.25$ as the rest frame radiation form to study the relation in a
general manner.

The Swift telescope payload is comprised of three instruments. Two of them
can detect high energy photons with one ranging from $0.3keV$ to $10keV$ and
the other from $15keV$ to $150keV$ (see
http://heasarc.gsfc.nasa.gov/docs/swift/about\_swift/). Here we study the
relation within a wide band ranging from $0.2keV$ to $1000keV$ which covers
both the Swift and BATSE bands. It should be noticed that, unlike the width
of pulses, the peak and total count depend strongly on the interval of
energy. We thus consider the following uniformly ranging channels (where $%
E_2=2E_1$): $[E_1,E_2]=[0.2,0.4]keV$ (channel A), $[0.5,1]keV$ (channel B), $%
[1,2]keV$ (channel C), $[2,4]keV$ (channel D), $[5,10]keV$ (channel E), $%
[10,20]keV$ (channel F), $[20,40]keV$ (channel G), $[50,100]keV$ (channel
H), $[100,200]keV$ (channel I), $[200,400]keV$ (channel J), and $%
[500,1000]keV$ (channel K). Note that, channel H is just the second channel
of BARSE.

Shown in Fig. 1 are the light curves of all these channels for the typical
hard and soft bursts, where we adopt the local Gaussian pulse with $\Delta
\tau _{\theta ,FWHM}=0.1$ and assign $\tau _{\theta ,\min }=0$ and $\tau
_{\theta ,0}=10\sigma +\tau _{\theta ,\min }$. We find in the figure that
while the profile of light curves is much less affected by the Lorentz
factor, in agreement with what revealed in Paper I, the relative magnitudes
and hence the relative total counts of different channel light curves depend
strongly on the latter. Listed in Table 1 are the peak count rate{\bf \ }$C_p
$ and the total count{\bf \ }$C_{total}$ of these channels associated with
different local pulse forms and widthes for the typical hard and soft
bursts, presented in the format relative to the corresponding values of
channel H. (Note that the absolute values would make sense only when all the
constants concerned such as $D$, $R_c$, and $I_0$ are available.) It shows
that relative values of these quantities depend obviously on the energy
concerned as well as the adopted Lorentz factor, but they are less affected
by the width and form of local pulses. Relations between the peak count rate
and energy and the total count and energy associated with these channels in
the case of adopting the local Gaussian pulse with $\Delta \tau _{\theta
,FWHM}=0.1$ for the typical hard and soft bursts are displayed in Fig. 2.
Shown in the figure, there exist semi-power law relationships between the
peak count rate and energy and between the total count and energy for both
the typical hard and soft bursts. The power law range for the hard burst
covers the BATSE band while that for the soft burst spans over a much wider
band, indicating that one could observe the power law relationship for the
soft burst but not the hard burst in the Swift band. It seems that, for a
bursts, the total count of a channel is proportional to the corresponding
peak count rate. Shown in Fig. 3 are the relation between the two quantities
for the typical hard and soft bursts in the case of adopting the local
Gaussian pulse with $\Delta \tau _{\theta ,FWHM}=0.1$. A semi-power law
relationship between the two quantities is observed for both bursts, where
the power law index for the hard burst is $1.34$ and that for the soft one
is $1.12$.

One might observe that, when adopting the channels defined above, some
counts such as those within $[0.4,0.5]keV$ would be missed. Following the
way that the BATSE team adopted, one can simply consider channels not
uniformly ranged but includes all the counts observed. Here, we consider
another set of channels which are similar to the previous ones but include
all counts. They are $[E_1,E_2]=[0.2,0.5]keV$ (channel Aa), $[0.5,1]keV$
(channel B), $[1,2]keV$ (channel C), $[2,5]keV$ (channel Da), $[5,10]keV$
(channel E), $[10,20]keV$ (channel F), $[20,50]keV$ (channel Ga), $%
[50,100]keV$ (channel H), $[100,300]keV$ (channel Ia), and $[300,1000]keV$
(channel Kb), where, the last four channels are just the BATSE channels.
Comparing this set of channels with the previous set, one would find that
channels Aa, Da, Ga, Ia, and Kb span over larger energy ranges than the
corresponding channels A, D, G, I, and K do, respectively. Presented in
Table 2 are the relative peak count rate and the relative total count of
channels Aa, Da, Ga, Ia, and Kb arising from the local Gaussian pulse with
different values of width for the typical hard and soft bursts. It shows
that, due to the larger energy ranges, the two quantities of channels Aa,
Da, Ga, Ia, and Kb are significantly larger than that of the corresponding
channels A, D, G, I, and K, respectively. Relations between the peak count
rate and energy and the total count and energy deduced from channels Aa, Da,
Ga, Ia, and Kb in the case of the local Gaussian pulse with $\Delta \tau
_{\theta ,FWHM}=0.1$ for the typical hard and soft bursts are also displayed
in Fig. 2. We find in the figure that the peak count rate and the total
count of the channels with larger energy ranges are obviously larger than
those of the smaller energy range channels, and the two quantities of the
former channels form by themselves continuous relationships with energy,
which are different from the relationships deduced from the smaller energy
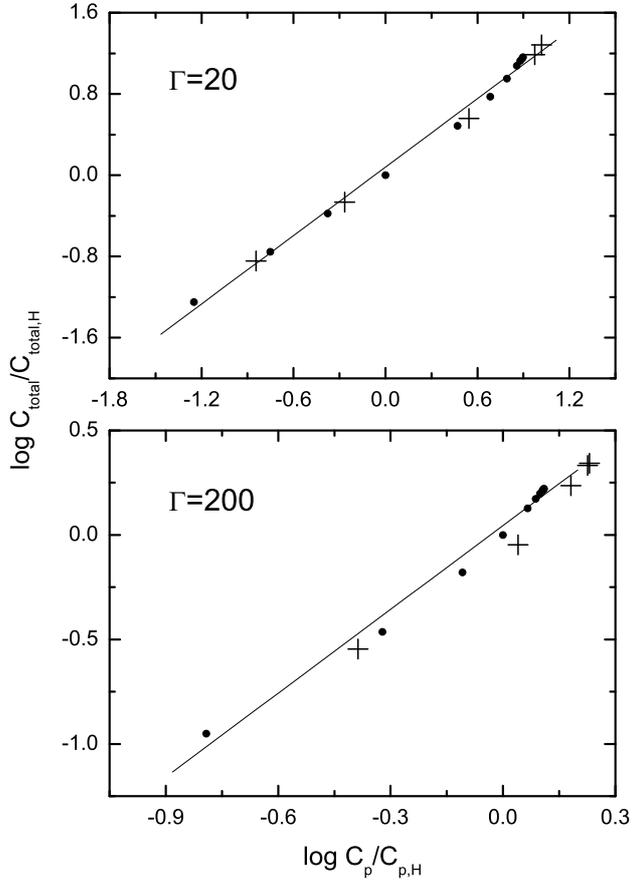
range channels. The relation between the peak count rate and total count of
channels Aa, Da, Ga, Ia, and Kb arising from the local Gaussian pulse with $%
\Delta \tau _{\theta ,FWHM}=0.1$ for the typical hard and soft bursts is
also shown in Fig. 3. The two quantities follow almost the same power law
relationships deduced from the uniformly ranging channels (channels A, B, C,
D, E, F, G, H, I, J, and K).

\subsection{The case of other rest frame radiation forms}

One could observe from Tables 1 and 2 that the pulse form and width do not
significantly affect the relative values of the peak count rate and total
count, while the Lorentz factor does. This suggests that, when performing a
statistical analysis on the relationships discussed above, the distribution
of the Lorentz factor would play an important role. We wonder if the rest
frame radiation form would also be an important factor in producing the
relationships. Let us study the relation in the cases of other rest frame
radiation forms, where only the case of the local Gaussian pulse with $%
\Delta \tau _{\theta ,FWHM}=0.1$ is considered.

Presented in Tables 3 and 4 (see Appendix B) are the relative values of the
peak count rate and total count of the light curves of various channels in
the case of local Gaussian pulses with $\Delta \tau _{\theta ,FWHM}=0.1$ for
two rest frame Band function spectra $(\alpha _0,\beta _0)=(0,-3.5)$ and $%
(\alpha _0,\beta _0)=(-1.5,-2)$, respectively. We find that the impacts of
the rest frame radiation form on the two quantities are very significant.
Displayed in Figs. 4 and 5 are the relationships between the two quantities
and energy for the two rest frame Band function spectra, respectively. We
find that, for a relatively steep rest frame spectrum, there is a turnover
in the relationships between the peak count rate and energy and between the
total count and energy, and the relationship between the peak count rate and
the total count exhibits a hook-like curve (see Fig. 4); for a relatively
flat rest frame spectrum, the power law relationships between the peak count
rate and energy and between the total count and energy would extend to the
lower energy range of the Swift band, and the relationship between the peak
count rate and the total count is a well-defined power law (see Fig. 5).
From these figures one can conclude that the values of{\bf \ }$C_p$ and $%
C_{total}$ depend obviously on the rest frame radiation form and the Lorentz
factor.

Besides the Band function, two other rest frame spectra are considered. One
is the thermal synchrotron spectrum which is written in the form $I_\nu
\propto (\nu /\nu _{0,s})\exp [-(\nu /\nu _{0,s})^{1/3}]$, where $\nu _{0,s}$
is a constant which includes all constants in the exponential index (Liang
et al. 1983). The other is the Comptonized spectrum which is written as $%
I_\nu \propto \nu ^{1+\alpha _{0,C}}\exp (-\nu /\nu _{0,C})$, where $\alpha
_{0,C}$ and $\nu _{0,C}$ are constants. Typical value $\alpha _{0,C}=-0.6$
(Schaefer et al. 1994) for the index of the Comptonized radiation will be
adopted. Corresponding to the typical hard and soft bursts, we take $\nu
_{0,s}=3.5\times 10^{-3}keVh^{-1}$ in the case of the rest frame thermal
synchrotron spectrum (see Qin 2002 Table 3), and take $\nu
_{0,C}=0.55keVh^{-1}$ in the case of the rest frame Comptonized spectrum
(see Qin 2002 Table 2). Relative values of the two quantities associated
with various channels for these two rest frame spectra are presented in
Tables 5 and 6 (see Appendix B), respectively. The values span several
orders of magnitudes. Shown in Fig. 6 are the relationships between the two
quantities and energy for the rest frame thermal synchrotron spectrum, which
is similar to Fig. 4 where a steep form of spectra is adopted. The most
obvious difference exhibits in the relationships between the peak count rate
and energy and between the total count and energy associated with the
typical soft burst, where the turnover is hardly detectable. The figure for
the rest frame Comptonized spectrum is quite similar to Fig. 6 and therefore
is omitted.

\subsection{The case of a rest frame spectrum varying with time}

Let us study the relation under the assumption that the rest frame radiation
form varies with time (this phenomenon is common in observation). With the
95 bursts fitted with the Band function in Preece et al. (2000), one could
find that the low and high energy power law indexes, $\alpha $ and $\beta $,
of the sources develop in the way $\alpha =-0.63-0.20(t-t_{\min })/(t_{\max
}-t_{\min })$ and $\beta =-2.44-0.42(t-t_{\min })/(t_{\max }-t_{\min })$ in
terms of statistics (see Qin 2004). We therefore consider an evolution of
the rest frame indexes $\alpha _0$ and $\beta _0$ following $\alpha
_0=-0.63-0.20(\tau _\theta -\tau _{\theta ,\min })/(\tau _{\theta ,\max
}-\tau _{\theta ,\min })$ and $\beta _0=-2.44-0.42(\tau _\theta -\tau
_{\theta ,\min })/(\tau _{\theta ,\max }-\tau _{\theta ,\min })$. Here, the
relation will be studied in the case of adopting the local Gaussian pulse
with $\Delta \tau _{\theta ,FWHM}=0.1$. We thus use $6\sigma $ to replace $%
\tau _{\theta ,\max }-\tau _{\theta ,\min }$ in these two relations. In the
same way, we take $\nu _{0,p}=0.75keVh^{-1}$ and assign $\Gamma =200$ to the
hard burst and $\Gamma =20$ to the soft one. Associated with the decreasing
of the indexes, we ignore the rising portion of the local pulse and assign
for the local Gaussian pulse that $\tau _{\theta ,0}=\tau _{\theta ,\min }$
and $\tau _{\theta ,\min }=0$. Listed in Table 7 (see Appendix B) are the
relative values of the peak count rate and total count of the light curves
of various channels in the case of local Gaussian pulses with $\Delta \tau
_{\theta ,FWHM}=0.1$ for the rest frame varying Band function spectrum. The
values also span several orders of magnitudes. The figure showing the
relationships between the two quantities and energy for the rest frame
varying Band function spectrum is also similar to Fig. 4 (it is therefore
omitted), suggesting that, creating the relation between different channel
light curves, the rest frame varying Band function acts like a steep form of
rest frame spectra.

\section{Direct comparison between count rates of different channel light
curves}

In this section we make a direct comparison between count rates of different
channel light curves, which may reveal other aspects of the relation between
them. All channels discussed above will be concerned.

\subsection{The case of a typical rest frame Band function spectrum}

Here we consider a rest frame Band function spectrum with $\alpha _0=-1$ and
$\beta _0=-2.25$.

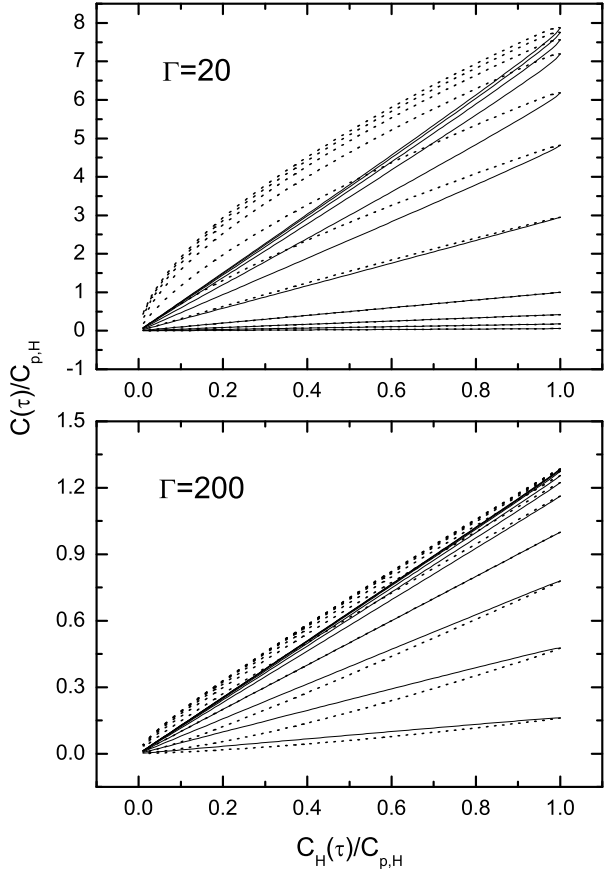
Presented in Fig. 7 are the curves of $C(\tau )$ versus $C_H(\tau )$ deduced
from the light curves of Fig.1, where $C_H(\tau )$ is the count rate of
channel H. This figure would be able to tell if the rising or decaying
speeds of the light curves of different channels are the same. We find that,
generally, the speeds are not the same, and for each channel, the rising and
decaying parts of $C(\tau )$ in the figure constitute a close curve. For the
typical soft burst, the speed of rising or decaying of the count rate of
higher energy bands is the same for different channels while the speed in
lower energy bands varies significantly. Corresponding to a certain value of
the count rate of channel H, there are two values of the count rate of lower
energy channels, where the one associated with the rising portion of the
light curve is larger than the one associated with the decaying phase. This
bi-valued character can also be seen in the case of the typical hard burst,
where for higher energy channels the one associated with the rising portion
of the light curve is smaller than the one associated with the decaying
phase. Besides the peak count rate discussed above, there are two characters
of the close curves in Fig. 7. One is the maximum value of the deviation of
the two values of the count rate of a channel corresponding to a certain
value of that of channel H. The other is the area confined by each close
curve in the figure. We define the maximum deviation of the two count rates
of a channel corresponding to a certain count rate of channel H as $\Delta
C_{\max }\equiv C_r/C_{p,H}-C_d/C_{p,H}$ when $|C_r/C_{p,H}-C_d/C_{p,H}|$ is
maximum, where $C_r$ and $C_d$ are the count rates of the rising portion and
decaying phase of the light curve, respectively, and $C_{p,H}$\ is the peak
count rate of channel H. In addition, we define an area $\Delta S$ confined
by the close curve as the area under the rising portion curve minus that
under the decaying phase curve in the plot of $C(\tau )$ versus $C_H(\tau )$%
, where all count rates are normalized to $C_{p,H}$ (see Fig. 7). According
to their definition, the values of both $\Delta C_{\max }$ and $\Delta S$
can be positive or negative. Presented in Table 8 are the values of $\Delta
C_{\max }$ and $\Delta S$ associated with different channel light curves in
the case of local Gaussian pulses with various values of $\Delta \tau
_{\theta ,FWHM}$ for the rest frame Band function spectrum with $\alpha _0=-1
$ and $\beta _0=-2.25$. We find that to determine these two quantities the
Lorentz factor plays an important role while the width of local pulses does
not. Displayed in Fig. 8 are the relationships between $\Delta C_{\max }$
and energy, and between $\Delta S$ and energy, and between $\Delta C_{\max }$
and $\Delta S$ themselves. It shows that, generally, both $\Delta C_{\max }$
and $\Delta S$ increase with the increasing of energy, and in the
relationships between $\Delta C_{\max }$ and $\log E/keV$ and between $%
\Delta S$ and $\log E/keV$, a turnover will be observed in the case of the
typical hard burst and a platform will be seen in the case of the typical
soft burst in a higher energy range. The two quantities are linearly
correlated. The slope for the typical hard burst is $0.70$ while that for
the typical soft burst is $0.72$.

\subsection{The case of other rest frame spectra}

The cases of other rest frame spectra discussed above are studied here,
including: two rest frame Band function spectra with $(\alpha _0,\beta
_0)=(0,-3.5)$ and $(\alpha _0,\beta _0)=(-1.5,-2)$, respectively; the
thermal synchrotron spectrum $I_\nu \propto (\nu /\nu _{0,s})\exp [-(\nu
/\nu _{0,s})^{1/3}]$; the Comptonized spectrum $I_\nu \propto \nu ^{1+\alpha
_{0,C}}\exp (-\nu /\nu _{0,C})$; and a varying Band function with its
indexes changing following $\alpha _0=-0.63-0.20(\tau _\theta -\tau _{\theta
,\min })/(\tau _{\theta ,\max }-\tau _{\theta ,\min })$ and $\beta
_0=-2.44-0.42(\tau _\theta -\tau _{\theta ,\min })/(\tau _{\theta ,\max
}-\tau _{\theta ,\min })$. We will take the same parameters adopted above to
calculate the two quantities $\Delta C_{\max }$ and $\Delta S$. Listed in
Tables 9-13 (see Appendix B) are the values of $\Delta C_{\max }$ and $%
\Delta S$ calculated in the case of the local Gaussian pulse with $\Delta
\tau _{\theta ,FWHM}=0.1$, adopting these rest frame spectra. It shows that,
to produce $\Delta C_{\max }$ and $\Delta S$, the Lorentz factor plays an
important role. It is interesting that, for the typical soft burst the
values of $\Delta C_{\max }$ and $\Delta S$ span over one order of
magnitudes, while for the typical hard burst they vary mildly. This suggests
that, for a certain radiation form, the ranges of $\Delta C_{\max }$ and $%
\Delta S$ are sensitive to the Lorentz factor. Shown in Figs. 9-13 are the
relationship between $\Delta C_{\max }$ and energy associated with these
rest frame spectra (the relationship between $\Delta S$ and energy is not
presented in these figures due to the similarity to the relationship between
$\Delta C_{\max }$ and energy). The linear relationship between $\Delta
C_{\max }$ and $\Delta S$ is almost the same for different cases discussed
here and therefore the corresponding figures are omitted. For a relatively
flat rest frame spectrum, $\Delta C_{\max }$ increases with energy, where a
platform in the high energy range can be observed (see Fig. 10). The
platform shifts to lower energy bands when the Lorentz factor becomes
smaller. For a relatively steep rest frame spectrum, $\Delta C_{\max }$
decreases with energy in lower bands while increases in higher bands. For
the typical soft burst, there is a platform in the higher bands, while for
the typical hard burst, the platform is replaced by a convex curve (see Fig.
9). Exhibited in the relationship, the behavior of the rest frame thermal
synchrotron, Comptonized and the varying Band function spectra is similar to
that of the rest frame steep spectrum (see Figs. 9 and 11-13). The values of
the two quantities rely obviously on the rest frame radiation form.

\section{Discussion and conclusions}

In this paper, the relation between different channel light curves of GRB
pulses is studied in aspects other than the width of pulses. The peak count
rate{\bf \ }$C_p$ and the total count{\bf \ }$C_{total}$ are employed. To
make a direct comparison between count rates of different channel light
curves we introduce a plot of $C(\tau )$ versus $C_H(\tau )$, where $C(\tau )
$ is the count rate of a channel and $C_H(\tau )$ is the count rate of
channel H. According to the plot we define $\Delta C_{\max }$ as the maximum
deviation of the two count rate values of $C(\tau )$ associated with a same
count rate value of $C_H(\tau )$ and define $\Delta S$ as the area confined
by the close curve of $C(\tau )$ in the plot to measure the difference of
the rising and decaying portions of a light curve relative to the count rate
of channel H. Predictions on the four quantities{\bf \ }$C_p$, $C_{total}$, $%
\Delta C_{\max }$, and $\Delta S$ within the Swift band, calculated with
different rest frame radiation forms and two typical Lorentz factors, are
made and presented (see the tables), which would make the test of our model
with the coming Swift data easier.

As shown in Fig. 7, a positive value of $\Delta C_{\max }$ (or $\Delta S$)
suggests that the ratio of $C(\tau )$ to $C_H(\tau )$ in the rising portion
of $C(\tau )$ is generally larger than that in the decaying phase, and a
negative value of $\Delta C_{\max }$ (or $\Delta S$) means the opposite.

The analysis shows that the four quantities{\bf \ }$C_p$, $C_{total}$, $%
\Delta C_{\max }$, and $\Delta S$ depend obviously on energy. The
relationships between them and energy are not significantly affected by the
local pulse form and width. Instead, they rely on the Lorentz factor and the
rest frame radiation form.

In the case of a relatively flat rest frame spectrum, both{\bf \ }$C_p$ and $%
C_{total}$ decrease with energy (see Fig. 5). As the radiation form is not
significantly affected by the Doppler effect of fireballs (see Qin 2002), we
suspect that this trend might be a consequence of the relatively large
amount of rest frame low energy photons which shift to X-ray or low
gamma-ray bands. In the case of a relatively steep rest frame spectrum,
there is a turnover in the relationship between $C_p$ (or $C_{total}$) and
energy (see Figs. 4 and 6). This might be caused by the position of the peak
energy $E_p$ of the observed spectrum. As the spectrum is steep, the number
of both higher and lower energy photons must be relatively small, and the
peak count rate would be observed in the channel around $E_p$. Indeed, we
find in Figs. 4 and 6 that the turnover appears in higher energy band when
the Lorentz factor is large (note that $E_p\propto \Gamma $; see, e.g., Qin
2002).

Shown in Figs. 9-13 we find that the value of $\Delta C_{\max }$ is negative
in low energy bands. This suggests that, compared with those of channel H,
both the increasing speed of the low energy photon number in the rising
phase and the decreasing speed of the number in the decaying phase are
smaller. We regard this as a consequence of the expanding fireball, where
high energy photons mainly come from the small area close to the line of
sight and hence the number of the photons increases rapidly while low energy
photons come from both the small area close to the line of sight and most of
the rest area of the fireball surface and hence the number of the photons
increases slowly, and for the same reason the number of high energy photons
decreases rapidly while the number of low energy photons decreases slowly in
the decaying phase.

\acknowledgments

This work was supported by the Special Funds for Major State Basic Research
Projects (``973'') and National Natural Science Foundation of China (No.
10273019).

\appendix

\section{Formulas used throughout this paper}

Formulas used throughout this paper can be found elsewhere (e.g., Qin 2004).
They are presented in the following so that it would be convenient to employ
them (e.g., when parameters of the formulas are concerned).

In this paper, we consider a highly symmetric and relativistically expanding
fireball emitting when (internal or external) shocks occur. As shown in Qin
(2004), the expected count rate of a fireball expanding with a Lorentz
factor $\Gamma >1$, measured within frequency interval [$\nu _1$, $\nu _2$],
can be determined by
\begin{equation}
C(\tau )=\frac{2\pi R_c^2}{hD^2}\frac{\int_{\widetilde{\tau
}_{\theta ,\min }}^{\widetilde{\tau }_{\theta ,\max
}}[\widetilde{I}(\tau _\theta
)(1+\beta \tau _\theta )^2(1-\tau +\tau _\theta )\int_{\nu _1}^{\nu _2}\frac{%
g_{0,\nu }(\nu _{0,\theta })}\nu d\nu ]d\tau _\theta }{\Gamma ^3(1-\beta
)^2(1+\frac \beta {1-\beta }\tau )^2},
\end{equation}
with $\tau _{\min }\leq \tau \leq \tau _{\max }$, $\tau _{\min }\equiv
(1-\beta )\tau _{\theta ,\min }$, $\tau _{\max }\equiv 1+\tau _{\theta ,\max
}$, $\tau \equiv (t-\frac Dc+\frac{R_c}c-t_c)/\frac{R_c}c$, and $\tau
_\theta \equiv (t_\theta -t_c)/\frac{R_c}c$, where $t$ is the observation
time measured by the distant observer, $t_\theta $ is the local time
measured by the local observer located at the place encountering the
expanding fireball surface at the position of $\theta $ relative to the
center of the fireball, $t_c$ is the initial local time, $R_c$ is the radius
of the fireball measured at $t_\theta =t_c$, $D$ is the distance from the
fireball to the observer, $\widetilde{I}(\tau _\theta )$ represents the
development of the intensity measured by the local observer, and $g_{0,\nu
}(\nu _{0,\theta })$ describes the rest frame radiation, and $\nu _{0,\theta
}=(1-\beta +\beta \tau )\Gamma \nu /(1+\beta \tau _\theta )$, $\widetilde{%
\tau }_{\theta ,\min }=\max \{\tau -1,\tau _{\theta ,\min }\}$, and $%
\widetilde{\tau }_{\theta ,\max }=\min \{\tau /(1-\beta ),\tau _{\theta
,\max }\}$, with $\tau _{\theta ,\min }$ and $\tau _{\theta ,\max }$ being
the upper and lower limits of $\tau _\theta $, respectively, which confine $%
\widetilde{I}(\tau _\theta )$. (Note that, since the limit of the Lorentz
factor is $\Gamma >1$, the formula can be applied in the cases of
relativistic, sub-relativistic, and non-relativistic motions.)

We consider in this paper two kinds of local pulses. The first is a local
Gaussian pulse. The intensity is assumed to be
\begin{equation}
\widetilde{I}(\tau _\theta )=I_0\exp [-(\frac{\tau _\theta -\tau
_{\theta ,0}}\sigma )^2]\qquad \qquad \qquad \qquad \qquad (\tau
_{\theta ,\min }\leq \tau _\theta ),
\end{equation}
where $I_0$, $\sigma $, $\tau _{\theta ,0}$ and $\tau _{\theta ,\min }$ are
constants. As shown in Paper I, there is a constraint to the lower limit of $%
\tau _\theta $. Due to this constraint, it is impossible to take a negative
infinity value of $\tau _{\theta ,\min }$ and therefore the interval between
$\tau _{\theta ,0}$ and $\tau _{\theta ,\min }$ must be limited. We hence
assign $\tau _{\theta ,0}=10\sigma +\tau _{\theta ,\min }$ so that the
interval between $\tau _{\theta ,0}$ and $\tau _{\theta ,\min }$ would be
large enough to make the rising part of the local pulse close to that of the
Gaussian pulse. The $FWHM$ width of the Gaussian pulse is $\Delta \tau
_{\theta ,FWHM}=2\sqrt{\ln 2}\sigma $, which leads to $\sigma =\Delta \tau
_{\theta ,FWHM}/2\sqrt{\ln 2}$. The second is a local pulse with a power law
rise and a power law decay, which is assumed to be
\begin{equation}
\widetilde{I}(\tau _\theta )=I_0\{
\begin{array}{c}
(\frac{\tau _\theta -\tau _{\theta ,\min }}{\tau _{\theta ,0}-\tau _{\theta
,\min }})^\mu \qquad \qquad \qquad \qquad \qquad (\tau _{\theta ,\min }\leq
\tau _\theta \leq \tau _{\theta ,0}) \\
(1-\frac{\tau _\theta -\tau _{\theta ,0}}{\tau _{\theta ,\max }-\tau
_{\theta ,0}})^\mu \qquad \qquad \qquad \qquad (\tau _{\theta ,0}<\tau
_\theta \leq \tau _{\theta ,\max })
\end{array}
,
\end{equation}
where $I_0$, $\mu $, $\tau _{\theta ,\min }$, $\tau _{\theta ,0}$ and $\tau
_{\theta ,\max }$ are constants. The peak of this intensity is at $\tau
_{\theta ,0}$, and the two $FWHM$ positions of this intensity before and
after $\tau _{\theta ,0}$ are $\tau _{\theta ,FWHM1}=2^{-1/\mu }\tau
_{\theta ,0}+(1-2^{-1/\mu })\tau _{\theta ,\min }$ and $\tau _{\theta
,FWHM2}=2^{-1/\mu }\tau _{\theta ,0}+(1-2^{-1/\mu })\tau _{\theta ,\max }$,
respectively. In the case of $\mu =2$, the $FWHM$ width of this local pulse
is $\Delta \tau _{\theta ,FWHM}=(1-1/\sqrt{2})(\tau _{\theta ,\max }-\tau
_{\theta ,\min })$, which leads to $\tau _{\theta ,\max }=\Delta \tau
_{\theta ,FWHM}/(1-1/\sqrt{2})+\tau _{\theta ,\min }$.

Through out this paper, we take $\tau _{\theta ,\min }=0$, and in the case
of the local power law pulse we take $\mu =2$ and assign $\tau _{\theta
,0}=\tau _{\theta ,\max }/2$ (in the case of the rest frame varying Band
function spectrum, we assign $\tau _{\theta ,0}=\tau _{\theta ,\min }$).

\section{Some tables}

Tables 3-7 and 9-13 are presented here in case of being useful when one is
able to check the predictions with these tables and observational data.

\clearpage

\begin{center}
Table 1. Relative peak count rates and total counts of the light curves of
Fig. 1

$
\begin{array}{c|c|cc|cc|cc|cc}
\hline\hline
&  & (Gau, & \Gamma =200) & (pow, & \Gamma =200) & (Gau, & \Gamma =20) &
(pow, & \Gamma =20) \\
\Delta \tau _{\theta ,F} & ch & \frac{C_p}{C_{p,H}} & \frac{C_{total}}{%
C_{total,H}} & \frac{C_p}{C_{p,H}} & \frac{C_{total}}{C_{total,H}} & \frac{%
C_p}{C_{p,H}} & \frac{C_{total}}{C_{total,H}} & \frac{C_p}{C_{p,H}} & \frac{%
C_{total}}{C_{total,H}} \\ \hline
0.01 &
\begin{array}{c}
A \\
B \\
C \\
D \\
E \\
F \\
G \\
H \\
I \\
J \\
K
\end{array}
&
\begin{array}{c}
1.27 \\
1.27 \\
1.27 \\
1.26 \\
1.24 \\
1.21 \\
1.16 \\
1.00 \\
0.787 \\
0.491 \\
0.168
\end{array}
&
\begin{array}{c}
1.68 \\
1.67 \\
1.66 \\
1.64 \\
1.59 \\
1.50 \\
1.34 \\
1.00 \\
0.662 \\
0.344 \\
0.112
\end{array}
&
\begin{array}{c}
1.27 \\
1.27 \\
1.27 \\
1.26 \\
1.24 \\
1.21 \\
1.16 \\
1.00 \\
0.786 \\
0.490 \\
0.168
\end{array}
&
\begin{array}{c}
1.75 \\
1.74 \\
1.73 \\
1.71 \\
1.64 \\
1.54 \\
1.36 \\
1.00 \\
0.659 \\
0.342 \\
0.111
\end{array}
&
\begin{array}{c}
7.51 \\
7.40 \\
7.22 \\
6.88 \\
5.95 \\
4.68 \\
2.92 \\
1.00 \\
0.420 \\
0.177 \\
0.0562
\end{array}
&
\begin{array}{c}
14.7 \\
14.2 \\
13.4 \\
12.0 \\
8.95 \\
5.92 \\
3.07 \\
1.00 \\
0.420 \\
0.177 \\
0.0562
\end{array}
&
\begin{array}{c}
7.54 \\
7.43 \\
7.25 \\
6.90 \\
5.97 \\
4.69 \\
2.92 \\
1.00 \\
0.420 \\
0.177 \\
0.0562
\end{array}
&
\begin{array}{c}
15.4 \\
14.7 \\
13.8 \\
12.2 \\
9.00 \\
5.93 \\
3.07 \\
1.00 \\
0.420 \\
0.177 \\
0.0562
\end{array}
\\ \hline
0.1 &
\begin{array}{c}
A \\
B \\
C \\
D \\
E \\
F \\
G \\
H \\
I \\
J \\
K
\end{array}
&
\begin{array}{c}
1.29 \\
1.28 \\
1.28 \\
1.27 \\
1.25 \\
1.22 \\
1.16 \\
1.00 \\
0.779 \\
0.477 \\
0.162
\end{array}
&
\begin{array}{c}
1.66 \\
1.66 \\
1.65 \\
1.63 \\
1.58 \\
1.49 \\
1.34 \\
1.00 \\
0.662 \\
0.344 \\
0.112
\end{array}
&
\begin{array}{c}
1.30 \\
1.29 \\
1.29 \\
1.28 \\
1.26 \\
1.23 \\
1.17 \\
1.00 \\
0.773 \\
0.467 \\
0.157
\end{array}
&
\begin{array}{c}
1.76 \\
1.75 \\
1.74 \\
1.72 \\
1.65 \\
1.54 \\
1.36 \\
1.00 \\
0.658 \\
0.341 \\
0.111
\end{array}
&
\begin{array}{c}
7.89 \\
7.78 \\
7.58 \\
7.21 \\
6.20 \\
4.83 \\
2.95 \\
1.00 \\
0.420 \\
0.177 \\
0.0562
\end{array}
&
\begin{array}{c}
14.6 \\
14.1 \\
13.3 \\
12.0 \\
8.94 \\
5.92 \\
3.07 \\
1.00 \\
0.420 \\
0.177 \\
0.0562
\end{array}
&
\begin{array}{c}
8.19 \\
8.06 \\
7.85 \\
7.45 \\
6.37 \\
4.93 \\
2.97 \\
1.00 \\
0.420 \\
0.177 \\
0.0562
\end{array}
&
\begin{array}{c}
15.4 \\
14.8 \\
13.9 \\
12.3 \\
9.00 \\
5.93 \\
3.07 \\
1.00 \\
0.420 \\
0.177 \\
0.0562
\end{array}
\\ \hline
1 &
\begin{array}{c}
A \\
B \\
C \\
D \\
E \\
F \\
G \\
H \\
I \\
J \\
K
\end{array}
&
\begin{array}{c}
1.30 \\
1.30 \\
1.30 \\
1.29 \\
1.27 \\
1.24 \\
1.17 \\
1.00 \\
0.770 \\
0.462 \\
0.155
\end{array}
&
\begin{array}{c}
1.76 \\
1.75 \\
1.74 \\
1.71 \\
1.64 \\
1.54 \\
1.36 \\
1.00 \\
0.658 \\
0.342 \\
0.111
\end{array}
&
\begin{array}{c}
1.35 \\
1.35 \\
1.34 \\
1.33 \\
1.31 \\
1.27 \\
1.20 \\
1.00 \\
0.747 \\
0.428 \\
0.141
\end{array}
&
\begin{array}{c}
1.78 \\
1.77 \\
1.76 \\
1.73 \\
1.66 \\
1.55 \\
1.36 \\
1.00 \\
0.658 \\
0.341 \\
0.111
\end{array}
&
\begin{array}{c}
8.33 \\
8.20 \\
7.98 \\
7.57 \\
6.46 \\
4.97 \\
2.98 \\
1.00 \\
0.420 \\
0.177 \\
0.0562
\end{array}
&
\begin{array}{c}
15.4 \\
14.8 \\
13.8 \\
12.3 \\
9.00 \\
5.93 \\
3.07 \\
1.00 \\
0.420 \\
0.177 \\
0.0562
\end{array}
&
\begin{array}{c}
9.45 \\
9.28 \\
9.00 \\
8.47 \\
7.09 \\
5.29 \\
3.03 \\
1.00 \\
0.420 \\
0.177 \\
0.0562
\end{array}
&
\begin{array}{c}
15.6 \\
14.9 \\
13.9 \\
12.3 \\
9.01 \\
5.93 \\
3.07 \\
1.00 \\
0.420 \\
0.177 \\
0.0562
\end{array}
\\ \hline\hline
\end{array}
$

Note: $\Delta \tau _{\theta ,F}$ --- $\Delta \tau _{\theta ,FWHM}$, $ch$ ---
channel, $Gau$ --- Gaussian pulse, $pow$ --- power law pulse.
\end{center}

\clearpage

\begin{center}
Table 2. Relative peak count rates and total counts of the light curves of
channels with larger energy ranges in the case of local Gaussian pulses\\[0pt%
]
$
\begin{array}{c|c|cc|cc}
\hline\hline
&  & (\Gamma = & 200) & (\Gamma = & 20) \\
\Delta \tau _{\theta ,F} & ch & \frac{C_p}{C_{p,H}} & \frac{C_{total}}{%
C_{total,H}} & \frac{C_p}{C_{p,H}} & \frac{C_{total}}{C_{total,H}} \\ \hline
0.01 &
\begin{array}{c}
Aa \\
Da \\
Ga \\
Ia \\
Kb
\end{array}
&
\begin{array}{c}
1.68 \\
1.67 \\
1.51 \\
1.11 \\
0.426
\end{array}
&
\begin{array}{c}
2.22 \\
2.17 \\
1.73 \\
0.895 \\
0.284
\end{array}
&
\begin{array}{c}
9.91 \\
8.98 \\
3.48 \\
0.542 \\
0.143
\end{array}
&
\begin{array}{c}
19.4 \\
15.5 \\
3.63 \\
0.542 \\
0.143
\end{array}
\\ \hline
0.1 &
\begin{array}{c}
Aa \\
Da \\
Ga \\
Ia \\
Kb
\end{array}
&
\begin{array}{c}
1.70 \\
1.68 \\
1.52 \\
1.10 \\
0.410
\end{array}
&
\begin{array}{c}
2.20 \\
2.15 \\
1.72 \\
0.896 \\
0.284
\end{array}
&
\begin{array}{c}
10.4 \\
9.40 \\
3.51 \\
0.542 \\
0.143
\end{array}
&
\begin{array}{c}
19.2 \\
15.4 \\
3.63 \\
0.542 \\
0.143
\end{array}
\\ \hline
1 &
\begin{array}{c}
Aa \\
Da \\
Ga \\
Ia \\
Kb
\end{array}
&
\begin{array}{c}
1.72 \\
1.70 \\
1.53 \\
1.08 \\
0.394
\end{array}
&
\begin{array}{c}
2.32 \\
2.26 \\
1.75 \\
0.891 \\
0.282
\end{array}
&
\begin{array}{c}
11.0 \\
9.86 \\
3.54 \\
0.542 \\
0.143
\end{array}
&
\begin{array}{c}
20.3 \\
15.7 \\
3.63 \\
0.542 \\
0.143
\end{array}
\\ \hline\hline
\end{array}
$
\end{center}

\clearpage

\begin{center}
Table 3. Relative peak count rates and total counts of the light curves of
various channels in the case of the local Gaussian pulse with $\Delta \tau
_{\theta ,FWHM}=0.1$ for the rest frame Band function spectrum with $\alpha
_0=0$ and $\beta _0=-3.5$ \\[0pt]
$
\begin{array}{c|cc|cc}
\hline\hline
& (\Gamma = & 200) & (\Gamma = & 20) \\
ch & \frac{C_p}{C_{p,H}} & \frac{C_{total}}{C_{total,H}} & \frac{C_p}{C_{p,H}%
} & \frac{C_{total}}{C_{total,H}} \\ \hline
A & 0.00675 & 0.0188 & 0.398 & 2.91 \\
B & 0.0168 & 0.0463 & 0.964 & 6.31 \\
C & 0.0335 & 0.0902 & 1.83 & 10.2 \\
D & 0.0662 & 0.172 & 3.29 & 14.4 \\
E & 0.160 & 0.372 & 6.01 & 17.0 \\
F & 0.304 & 0.601 & 7.19 & 13.9 \\
G & 0.547 & 0.848 & 5.33 & 6.94 \\
H & 1.00 & 1.00 & 1.00 & 1.00 \\
I & 1.20 & 0.815 & 0.177 & 0.177 \\
J & 0.887 & 0.408 & 0.0313 & 0.0313 \\
K & 0.167 & 0.0588 & 0.00316 & 0.00316 \\
Aa & 0.0101 & 0.0282 & 0.595 & 4.30 \\
Da & 0.0990 & 0.253 & 4.77 & 19.8 \\
Ga & 0.793 & 1.16 & 6.20 & 7.86 \\
Ia & 1.79 & 1.11 & 0.201 & 0.202 \\
Kb & 0.608 & 0.231 & 0.0131 & 0.0132 \\ \hline\hline
\end{array}
$
\end{center}

\clearpage

\begin{center}
Table 4. Relative peak count rates and total counts of the light curves of
various channels in the case of the local Gaussian pulse with $\Delta \tau
_{\theta ,FWHM}=0.1$ for the rest frame Band function spectrum with $\alpha
_0=-1.5$ and $\beta _0=-2$ \\[0pt]
$
\begin{array}{c|cc|cc}
\hline\hline
& (\Gamma = & 200) & (\Gamma = & 20) \\
ch & \frac{C_p}{C_{p,H}} & \frac{C_{total}}{C_{total,H}} & \frac{C_p}{C_{p,H}%
} & \frac{C_{total}}{C_{total,H}} \\ \hline
A & 17.9 & 20.1 & 40.2 & 51.4 \\
B & 11.3 & 12.7 & 25.3 & 32.0 \\
C & 7.98 & 8.96 & 17.6 & 22.0 \\
D & 5.63 & 6.30 & 12.2 & 14.8 \\
E & 3.53 & 3.92 & 7.15 & 8.16 \\
F & 2.47 & 2.70 & 4.47 & 4.75 \\
G & 1.70 & 1.81 & 2.48 & 2.50 \\
H & 1.00 & 1.00 & 1.00 & 1.00 \\
I & 0.625 & 0.582 & 0.500 & 0.500 \\
J & 0.348 & 0.306 & 0.250 & 0.250 \\
K & 0.140 & 0.123 & 0.100 & 0.100 \\
Aa & 22.4 & 25.3 & 50.5 & 64.6 \\
Da & 7.06 & 7.91 & 15.2 & 18.4 \\
Ga & 2.12 & 2.26 & 2.98 & 3.00 \\
Ia & 0.857 & 0.789 & 0.667 & 0.669 \\
Kb & 0.327 & 0.287 & 0.233 & 0.235 \\ \hline\hline
\end{array}
$
\end{center}

\clearpage

\begin{center}
Table 5. Relative peak count rates and total counts of the light curves of
various channels in the case of the local Gaussian pulse with $\Delta \tau
_{\theta ,FWHM}=0.1$ for the rest frame thermal synchrotron spectrum

$
\begin{array}{c|cc|cc}
\hline\hline
& (\Gamma = & 200) & (\Gamma = & 20) \\
ch & \frac{C_p}{C_{p,H}} & \frac{C_{total}}{C_{total,H}} & \frac{C_p}{C_{p,H}%
} & \frac{C_{total}}{C_{total,H}} \\ \hline
A & 0.0968 & 0.267 & 3.53 & 12.7 \\
B & 0.195 & 0.485 & 5.54 & 16.8 \\
C & 0.315 & 0.711 & 7.02 & 18.3 \\
D & 0.481 & 0.972 & 7.90 & 17.6 \\
E & 0.757 & 1.28 & 7.33 & 13.1 \\
F & 0.958 & 1.40 & 5.53 & 8.28 \\
G & 1.08 & 1.35 & 3.26 & 4.10 \\
H & 1.00 & 1.00 & 1.00 & 1.00 \\
I & 0.755 & 0.635 & 0.257 & 0.217 \\
J & 0.445 & 0.314 & 0.0397 & 0.0283 \\
K & 0.137 & 0.0766 & 0.00125 & 0.000719 \\
Aa & 0.141 & 0.384 & 4.98 & 17.5 \\
Da & 0.679 & 1.34 & 10.5 & 22.7 \\
Ga & 1.43 & 1.74 & 3.88 & 4.80 \\
Ia & 1.05 & 0.854 & 0.289 & 0.242 \\
Kb & 0.370 & 0.229 & 0.0105 & 0.00687 \\ \hline\hline
\end{array}
$
\end{center}

\clearpage

\begin{center}
Table 6. Relative peak count rates and total counts of the light curves of
various channels in the case of the local Gaussian pulse with $\Delta \tau
_{\theta ,FWHM}=0.1$ for the rest frame Comptonized spectrum\\[0pt]
$
\begin{array}{c|cc|cc}
\hline\hline
& (\Gamma = & 200) & (\Gamma = & 20) \\
ch & \frac{C_p}{C_{p,H}} & \frac{C_{total}}{C_{total,H}} & \frac{C_p}{C_{p,H}%
} & \frac{C_{total}}{C_{total,H}} \\ \hline
A & 0.156 & 0.263 & 2.88 & 13.4 \\
B & 0.224 & 0.377 & 4.07 & 18.0 \\
C & 0.295 & 0.492 & 5.19 & 21.3 \\
D & 0.386 & 0.634 & 6.38 & 23.3 \\
E & 0.545 & 0.853 & 7.47 & 21.1 \\
F & 0.695 & 1.01 & 6.98 & 15.2 \\
G & 0.854 & 1.10 & 4.70 & 7.48 \\
H & 1.00 & 1.00 & 1.00 & 1.00 \\
I & 0.935 & 0.720 & 0.0708 & 0.0494 \\
J & 0.629 & 0.354 & 0.000442 & 0.000218 \\
K & 0.134 & 0.0473 & 2.15\times 10^{-10} & 7.35\times 10^{-11} \\
Aa & 0.216 & 0.365 & 3.99 & 18.4 \\
Da & 0.534 & 0.872 & 8.66 & 30.8 \\
Ga & 1.16 & 1.46 & 5.53 & 8.53 \\
Ia & 1.35 & 0.972 & 0.0713 & 0.0503 \\
Kb & 0.457 & 0.202 & 3.26\times 10^{-6} & 1.39\times 10^{-6} \\ \hline\hline
\end{array}
$
\end{center}

\clearpage

\begin{center}
Table 7. Relative peak count rates and total counts of the light curves of
various channels in the case of the local Gaussian pulse with $\Delta \tau
_{\theta ,FWHM}=0.1$ for the rest frame varying Band function spectrum\\[0pt]
$
\begin{array}{c|cc|cc}
\hline\hline
& (\Gamma = & 200) & (\Gamma = & 20) \\
ch & \frac{C_p}{C_{p,H}} & \frac{C_{total}}{C_{total,H}} & \frac{C_p}{C_{p,H}%
} & \frac{C_{total}}{C_{total,H}} \\ \hline
A & 0.204 & 0.351 & 2.59 & 7.45 \\
B & 0.281 & 0.480 & 3.50 & 9.52 \\
C & 0.357 & 0.603 & 4.31 & 10.8 \\
D & 0.451 & 0.748 & 5.13 & 11.4 \\
E & 0.609 & 0.954 & 5.74 & 9.99 \\
F & 0.750 & 1.08 & 5.20 & 7.14 \\
G & 0.893 & 1.14 & 3.40 & 3.68 \\
H & 1.00 & 1.00 & 1.00 & 1.00 \\
I & 0.906 & 0.714 & 0.359 & 0.358 \\
J & 0.592 & 0.368 & 0.129 & 0.129 \\
K & 0.174 & 0.100 & 0.0333 & 0.0331 \\
Aa & 0.281 & 0.484 & 3.56 & 10.2 \\
Da & 0.621 & 1.02 & 6.92 & 14.9 \\
Ga & 1.20 & 1.50 & 4.01 & 4.30 \\
Ia & 1.30 & 0.971 & 0.450 & 0.451 \\
Kb & 0.477 & 0.278 & 0.0920 & 0.0921 \\ \hline\hline
\end{array}
$
\end{center}

\clearpage

\begin{center}
Table 8. Values of $\Delta C_{\max }$ and $\Delta S$ in the case of local
Gaussian pulses for the rest frame Band function spectrum with $\alpha _0=-1$
and $\beta _0=-2.25$

$
\begin{array}{c|c|cc|cc}
\hline\hline
&  & (\Gamma = & 200) & (\Gamma = & 20) \\
\Delta \tau _{\theta ,F} & ch & \Delta C_{\max } & \Delta S & \Delta C_{\max
} & \Delta S \\ \hline
0.01 &
\begin{array}{c}
A \\
B \\
C \\
D \\
E \\
F \\
G \\
H \\
I \\
J \\
K \\
Aa \\
Da \\
Ga \\
Ia \\
Kb
\end{array}
&
\begin{array}{c}
-0.07 \\
-0.07 \\
-0.07 \\
-0.07 \\
-0.06 \\
-0.05 \\
-0.04 \\
0.00 \\
0.04 \\
0.05 \\
0.02 \\
-0.09 \\
-0.09 \\
-0.04 \\
0.07 \\
0.06
\end{array}
&
\begin{array}{c}
-0.05 \\
-0.05 \\
-0.05 \\
-0.04 \\
-0.04 \\
-0.03 \\
-0.02 \\
0.00 \\
0.02 \\
0.04 \\
0.02 \\
-0.06 \\
-0.06 \\
-0.03 \\
0.05 \\
0.04
\end{array}
&
\begin{array}{c}
-1.34 \\
-1.29 \\
-1.23 \\
-1.10 \\
-0.78 \\
-0.42 \\
-0.08 \\
0.00 \\
0.00 \\
0.00 \\
0.00 \\
-1.76 \\
-1.41 \\
-0.08 \\
0.00 \\
0.00
\end{array}
&
\begin{array}{c}
-0.91 \\
-0.88 \\
-0.84 \\
-0.75 \\
-0.53 \\
-0.28 \\
-0.05 \\
0.00 \\
0.00 \\
0.00 \\
0.00 \\
-1.21 \\
-0.96 \\
-0.05 \\
0.00 \\
0.00
\end{array}
\\ \hline
0.1 &
\begin{array}{c}
A \\
B \\
C \\
D \\
E \\
F \\
G \\
H \\
I \\
J \\
K \\
Aa \\
Da \\
Ga \\
Ia \\
Kb
\end{array}
&
\begin{array}{c}
-0.08 \\
-0.08 \\
-0.07 \\
-0.07 \\
-0.07 \\
-0.06 \\
-0.04 \\
0.00 \\
0.04 \\
0.06 \\
0.02 \\
-0.10 \\
-0.10 \\
-0.05 \\
0.08 \\
0.06
\end{array}
&
\begin{array}{c}
-0.05 \\
-0.05 \\
-0.05 \\
-0.05 \\
-0.05 \\
-0.04 \\
-0.03 \\
0.00 \\
0.03 \\
0.04 \\
0.02 \\
-0.07 \\
-0.07 \\
-0.03 \\
0.05 \\
0.04
\end{array}
&
\begin{array}{c}
-1.51 \\
-1.47 \\
-1.39 \\
-1.24 \\
-0.88 \\
-0.46 \\
-0.08 \\
0.00 \\
0.00 \\
0.00 \\
0.00 \\
-2.00 \\
-1.59 \\
-0.08 \\
0.00 \\
0.00
\end{array}
&
\begin{array}{c}
-1.09 \\
-1.05 \\
-1.00 \\
-0.89 \\
-0.63 \\
-0.33 \\
-0.05 \\
0.00 \\
0.00 \\
0.00 \\
0.00 \\
-1.44 \\
-1.14 \\
-0.05 \\
0.00 \\
0.00
\end{array}
\\ \hline
1 &
\begin{array}{c}
A \\
B \\
C \\
D \\
E \\
F \\
G \\
H \\
I \\
J \\
K \\
Aa \\
Da \\
Ga \\
Ia \\
Kb
\end{array}
&
\begin{array}{c}
-0.08 \\
-0.08 \\
-0.08 \\
-0.08 \\
-0.07 \\
-0.06 \\
-0.04 \\
0.00 \\
0.04 \\
0.06 \\
0.02 \\
-0.11 \\
-0.11 \\
-0.05 \\
0.08 \\
0.06
\end{array}
&
\begin{array}{c}
-0.06 \\
-0.06 \\
-0.06 \\
-0.06 \\
-0.05 \\
-0.05 \\
-0.03 \\
0.00 \\
0.03 \\
0.04 \\
0.02 \\
-0.08 \\
-0.08 \\
-0.04 \\
0.06 \\
0.04
\end{array}
&
\begin{array}{c}
-1.71 \\
-1.66 \\
-1.57 \\
-1.40 \\
-0.98 \\
-0.51 \\
-0.08 \\
0.00 \\
0.00 \\
0.00 \\
0.00 \\
-2.26 \\
-1.79 \\
-0.08 \\
0.00 \\
0.00
\end{array}
&
\begin{array}{c}
-1.28 \\
-1.24 \\
-1.17 \\
-1.04 \\
-0.72 \\
-0.37 \\
-0.06 \\
0.00 \\
0.00 \\
0.00 \\
0.00 \\
-1.69 \\
-1.33 \\
-0.06 \\
0.00 \\
0.00
\end{array}
\\ \hline\hline
\end{array}
$
\end{center}

\clearpage

\begin{center}
Table 9. Values of $\Delta C_{\max }$ and $\Delta S$ in the case of the
local Gaussian pulse with $\Delta \tau _{\theta ,FWHM}=0.1$ for the rest
frame Band function spectrum with $\alpha _0=0$ and $\beta _0=-3.5$\\[0pt]
$
\begin{array}{c|cc|cc}
\hline\hline
& (\Gamma = & 200) & (\Gamma = & 20) \\
ch & \Delta C_{\max } & \Delta S & \Delta C_{\max } & \Delta S \\ \hline
A & 0.00 & 0.00 & -0.21 & -0.15 \\
B & 0.00 & 0.00 & -0.50 & -0.35 \\
C & -0.01 & 0.00 & -0.92 & -0.64 \\
D & -0.01 & -0.01 & -1.55 & -1.10 \\
E & -0.03 & -0.02 & -2.36 & -1.70 \\
F & -0.05 & -0.03 & -2.11 & -1.53 \\
G & -0.06 & -0.04 & -0.81 & -0.58 \\
H & 0.00 & 0.00 & 0.00 & 0.00 \\
I & 0.16 & 0.11 & 0.00 & 0.00 \\
J & 0.26 & 0.17 & 0.00 & 0.00 \\
K & 0.07 & 0.05 & 0.00 & 0.00 \\
Aa & 0.00 & 0.00 & -0.31 & -0.22 \\
Da & -0.02 & -0.01 & -2.20 & -1.56 \\
Ga & -0.08 & -0.05 & -0.83 & -0.59 \\
Ia & 0.31 & 0.21 & 0.00 & 0.00 \\
Kb & 0.23 & 0.16 & 0.00 & 0.00 \\ \hline\hline
\end{array}
$
\end{center}

\clearpage

\begin{center}
Table 10. Values of $\Delta C_{\max }$ and $\Delta S$ in the case of the
local Gaussian pulse with $\Delta \tau _{\theta ,FWHM}=0.1$ for the rest
frame Band function spectrum with $\alpha _0=-1.5$ and $\beta _0=-2$\\[0pt]
$
\begin{array}{c|cc|cc}
\hline\hline
& (\Gamma = & 200) & (\Gamma = & 20) \\
ch & \Delta C_{\max } & \Delta S & \Delta C_{\max } & \Delta S \\ \hline
A & -0.42 & -0.30 & -2.96 & -2.13 \\
B & -0.27 & -0.19 & -1.83 & -1.31 \\
C & -0.19 & -0.13 & -1.23 & -0.88 \\
D & -0.13 & -0.09 & -0.79 & -0.57 \\
E & -0.08 & -0.05 & -0.37 & -0.26 \\
F & -0.05 & -0.03 & -0.14 & -0.10 \\
G & -0.02 & -0.02 & -0.01 & -0.01 \\
H & 0.00 & 0.00 & 0.00 & 0.00 \\
I & 0.01 & 0.01 & 0.00 & 0.00 \\
J & 0.02 & 0.01 & 0.00 & 0.00 \\
K & 0.01 & 0.01 & 0.00 & 0.00 \\
Aa & -0.53 & -0.37 & -3.71 & -2.66 \\
Da & -0.16 & -0.11 & -0.98 & -0.70 \\
Ga & -0.03 & -0.02 & -0.01 & -0.01 \\
Ia & 0.03 & 0.02 & 0.00 & 0.00 \\
Kb & 0.02 & 0.01 & 0.00 & 0.00 \\ \hline\hline
\end{array}
$
\end{center}

\clearpage

\begin{center}
Table 11. Values of $\Delta C_{\max }$ and $\Delta S$ in the case of the
local Gaussian pulse with $\Delta \tau _{\theta ,FWHM}=0.1$ for the rest
frame thermal synchrotron spectrum\\[0pt]
$
\begin{array}{c|cc|cc}
\hline\hline
& (\Gamma = & 200) & (\Gamma = & 20) \\
ch & \Delta C_{\max } & \Delta S & \Delta C_{\max } & \Delta S \\ \hline
A & -0.02 & -0.02 & -1.29 & -0.92 \\
B & -0.05 & -0.03 & -1.82 & -1.30 \\
C & -0.07 & -0.05 & -2.07 & -1.49 \\
D & -0.09 & -0.06 & -2.03 & -1.46 \\
E & -0.11 & -0.08 & -1.44 & -1.05 \\
F & -0.10 & -0.07 & -0.80 & -0.58 \\
G & -0.07 & -0.05 & -0.28 & -0.20 \\
H & 0.00 & 0.00 & 0.00 & 0.00 \\
I & 0.04 & 0.03 & 0.02 & 0.01 \\
J & 0.05 & 0.04 & 0.01 & 0.00 \\
K & 0.03 & 0.02 & 0.00 & 0.00 \\
Aa & -0.04 & -0.02 & -1.79 & -1.27 \\
Da & -0.12 & -0.09 & -2.61 & -1.89 \\
Ga & -0.08 & -0.06 & -0.31 & -0.22 \\
Ia & 0.07 & 0.05 & 0.02 & 0.02 \\
Kb & 0.06 & 0.04 & 0.00 & 0.00 \\ \hline\hline
\end{array}
$
\end{center}

\clearpage

\begin{center}
Table 12. Values of $\Delta C_{\max }$ and $\Delta S$ in the case of the
local Gaussian pulse with $\Delta \tau _{\theta ,FWHM}=0.1$ for the rest
frame Comptonized spectrum\\[0pt]
$
\begin{array}{c|cc|cc}
\hline\hline
& (\Gamma = & 200) & (\Gamma = & 20) \\
ch & \Delta C_{\max } & \Delta S & \Delta C_{\max } & \Delta S \\ \hline
A & -0.02 & -0.01 & -1.25 & -0.89 \\
B & -0.02 & -0.02 & -1.75 & -1.24 \\
C & -0.03 & -0.02 & -2.18 & -1.55 \\
D & -0.04 & -0.03 & -2.59 & -1.84 \\
E & -0.05 & -0.03 & -2.71 & -1.95 \\
F & -0.05 & -0.04 & -2.09 & -1.52 \\
G & -0.05 & -0.03 & -0.94 & -0.69 \\
H & 0.00 & 0.00 & 0.00 & 0.00 \\
I & 0.07 & 0.05 & 0.01 & 0.01 \\
J & 0.12 & 0.08 & 0.00 & 0.00 \\
K & 0.05 & 0.04 & 0.00 & 0.00 \\
Aa & -0.02 & -0.01 & -1.73 & -1.22 \\
Da & -0.05 & -0.04 & -3.48 & -2.48 \\
Ga & -0.06 & -0.04 & -1.02 & -0.75 \\
Ia & 0.14 & 0.10 & 0.01 & 0.01 \\
Kb & 0.13 & 0.09 & 0.00 & 0.00 \\ \hline\hline
\end{array}
$
\end{center}

\clearpage

\begin{center}
Table 13. Values of $\Delta C_{\max }$ and $\Delta S$ in the case of the
local Gaussian pulse with $\Delta \tau _{\theta ,FWHM}=0.1$ for the rest
frame varying Band function spectrum\\[0pt]
$
\begin{array}{c|cc|cc}
\hline\hline
& (\Gamma = & 200) & (\Gamma = & 20) \\
ch & \Delta C_{\max } & \Delta S & \Delta C_{\max } & \Delta S \\ \hline
A & -0.03 & -0.02 & -0.82 & -0.62 \\
B & -0.03 & -0.03 & -1.08 & -0.81 \\
C & -0.04 & -0.03 & -1.28 & -0.95 \\
D & -0.05 & -0.04 & -1.42 & -1.05 \\
E & -0.06 & -0.04 & -1.31 & -0.96 \\
F & -0.06 & -0.05 & -0.84 & -0.61 \\
G & -0.05 & -0.04 & -0.21 & -0.14 \\
H & 0.00 & 0.00 & 0.00 & 0.00 \\
I & 0.07 & 0.05 & 0.00 & 0.00 \\
J & 0.11 & 0.08 & 0.00 & 0.00 \\
K & 0.04 & 0.03 & 0.00 & 0.00 \\
Aa & -0.03 & -0.03 & -1.13 & -0.85 \\
Da & -0.07 & -0.05 & -1.89 & -1.40 \\
Ga & -0.06 & -0.05 & -0.21 & -0.14 \\
Ia & 0.14 & 0.10 & 0.01 & 0.00 \\
Kb & 0.11 & 0.08 & 0.00 & 0.00 \\ \hline\hline
\end{array}
$
\end{center}

\end{document}